# The TRAMOS pixel as a photo-detection device: design, architecture and building blocks

Nicolas T. Fourches [a1], Vishant Kumar[b], Yves Serruys [c], G. Gutierrez [c], F. Leprêtre [c], F. Jomard[d]

[a]*CEA IRFU/DEDIP, Université Paris-Saclay CEA Saclay 91191 Gif/Yvette, France*
[b]*CIMAP,14000, Caen France*
[c]*CEA/DEN/Jannus,Université Paris-Saclay CEA Saclay,91191 Gif/Yvette,France*
[d]*GEMAC , Université de Versailles-Saint Quentin en Yvelines , 55 avenue de Paris, 7800 Versailles, France*

**Abstract:**
The deep trapping gate device concept for charged particle detection was recently introduced in Saclay/IRFU. It is based on an n-MOS structure in which a buried gate, located below the n-channel, collects carriers which are generated by ionizing particles. They deposit their energy in a volume which extends in the bulk, below the buried gate. The n-channel device is based on holes in-buried gate localization. Source-drain current modulation occurs, measurable during readout. The buried gate (Deep Trapping Gate or DTG) contains deep level centers which can be introduced during process or may be made with a Quantum Well. The device can be scaled down providing a micron range resolution. The proof of principle for such a device was verified using 2D device and process simulations. Work under way focusses on the study of building blocks. In this contribution, the pixel proof of design, using existing fabrication techniques will be discussed first. The use of this pixel for photon imaging will be discussed.

---
[1]Corresponding Author. Nicolas Fourches: Tel.: +33(0)164463616; fax: +33(0)164463616: nicolas.fourches@cea.fr

## 1. TRAMOS pixel operation principle and background

### 1.1. Introduction: general concept and background

In 2010 the deep trapping gate device concept (or TRAMOS: TRApping MOS) for charged particle detection was introduced by Saclay/IRFU [1]. It is based on an n-MOS structure in which a buried gate is located below the n-channel. Particles that generate electron-hole pairs by depositing energy in the volume located below the channel can be detected in the following way. Generated holes drift towards the channel and become localized in the buried gate. The subsequent change in the buried gate charge state leads to a modulation of the source-drain current. This new buried gate contains localized deep level centers (Deep Trapping Gate or DTG) which can be introduced during process, or alternatively a SiGe quantum well (Valence Band Well) can be used. The buried gate is then effective in localizing charge carriers during an amount of time sufficiently long to ensure adequate pixel readout (Fig.1) The point to point resolution of such a device, reduced to one single transistor ($1\mu m^2$), follows the down-scaling rules of standard CMOS processes. This low area sets the resolution below 1 μm which is a significant improvement with respect to DEPFET [2] and CMOS pixels. The TRAMOS proof of principle was verified using (TCAD) 2D device and process simulations [3]. Because it is depleted during detection the TRAMOS is potentially harder with respect to bulk damage than previous CMOS pixels. The pixel proof of design, using existing fabrications technique will be presented here. The use of this pixel for photons detection and imaging both of low and high energy is also possible. These pixel-arrays are primarily intended to be used in the inner-detectors for vertex determination in high energy physics experiments. We will focus on the technological steps that are necessary to fabricate the elementary pixel.

### 1.2. Technological options

There are two designs for the TRAMOS concept, one based on deep traps (substitutional Zn) deliberately introduced in the n-channel device as a buried gate using ion implantation. This requires the control of the Zn concentration profile in order to get an effective trapping gate with no contamination of the rest of the device. Up to now simulations have proved the principle valid as far as the buried gate remains effective. These results have been published in 2010 [1]. Recent attempts to make a buried gate using ion implantation at high energies (1MeV) have shown that a Zn enriched gate may be created but that Zn exo-diffusion may be a cause for concern as Zn has a background density in the whole sample observed using SIMS and RBS. Further investigations should be made to fix this as the measurements have been made many weeks after irradiation with no thermal annealing performed. RTA (Rapid Thermal Anneal) after Zn implantation should be considered as an add-on to the process useful to possibly alleviate room temperature Zn migration. The other concept is from the inventor point of view the most promising. First it is based on the Quantum Well properties (Fig.2) For a SiGe 20 nm layer buried into silicon (Si/SiGe/Si) this gives a band diagram with a VB

(valence band) well (in the SiGe) and a CB (conduction band) barrier, if some growth conditions are met, in order to obtain a compressively strained layer. This kind of structure can be fabricated with a UHV/CVD technique although we have used ion-implantation as an alternative.

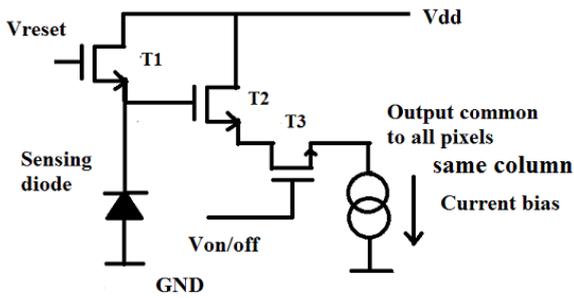

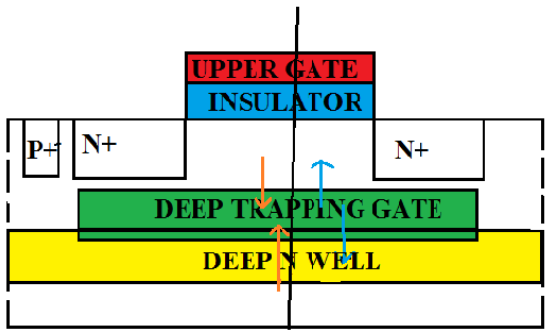

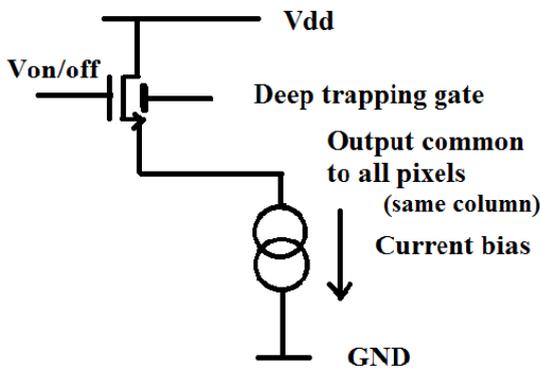

Fig. 1: Migration from a 3T pixel to a 1T design with reduced dimensions. The photo-generated carriers drift towards the upper gate (holes with a –Vb negative bias) and the substrate (electrons with backside grounded). In readout mode we use opposite bias to switch the MOSFET on.

The ion implantation can be made at high energy (1MeV) to obtain a deep Ge layer. Using an etching scheme and some Si epitaxy it would be possible to obtain an abrupt heterojunction as shown in (Fig. 3.).

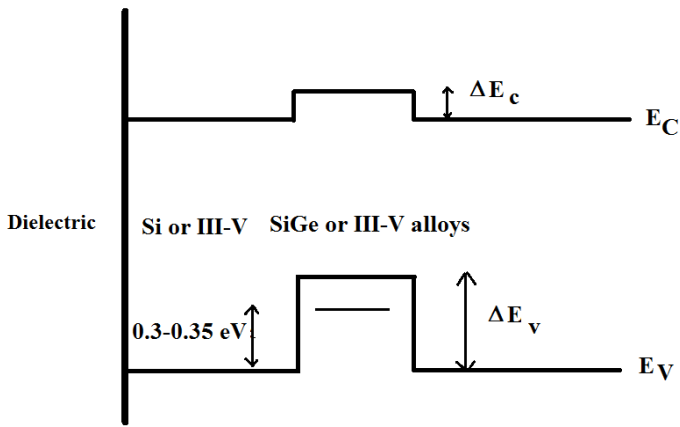

Fig.2: Quantum Well structure proposed for the pixel device. Alternatively III-V materials may be used. In these case the dielectric if necessary can be any suitable insulator. For Silicon based devices, thermal oxide can be partly replaced by other dielectrics such as CVD $HfO_2$.

In fact some comparable structures have been proposed and made as early as 1997 for the development of SiGe MODFETs (MOdulation Doped Field Effect Transistors) [4]. The fabrication techniques are similar to the ones necessary for the QW TRAMOS making the development of the TRAMOS close to our reach.

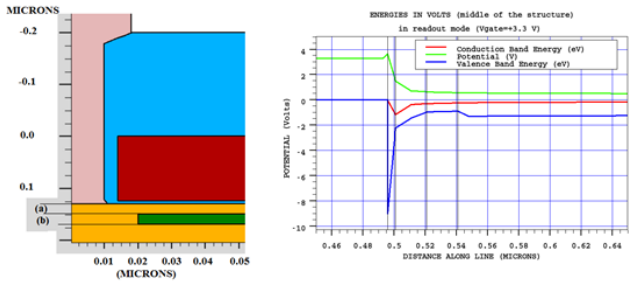

Fig. 3: Cut view of the structure used for simulations (left), with the band diagram of the structure (VB, CB, potential) .These results are from TCAD Silvaco simulations of the structure.

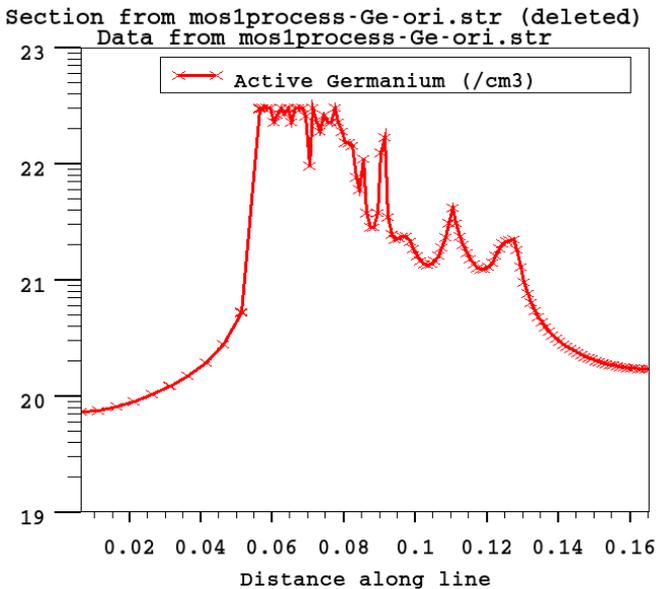

Fig.4: Simulated Ge profile in an implanted (50 keV) and etched (50 nm) and epitaxial Si grown (850 °C 50 min). This leads to an abrupt hetero-junction ($<<$ 5 nm).The diffusion coefficients are taken from [5].The noisy plot results on the left are not significant.

1.3. Simulation studies

In order to reduce development time, we have tried to replace a classical approach of test and trial by a procedure based on device and process simulations. A standard TCAD simulation package is used for this purpose. We can summarized the results in Table 1. Note that the characteristics of the pixel and especially its latency (defined here as the maximum time after a particle hit the pixel to obtain a significant signal for a correct readout) is strongly depend on the proportion of Ge in the SiGe layer when a SiGe QW is used.

Table 1: simulated characteristics of a typical pixel

| CVF (nA/V) | 20 nA/carrier | 60 µV/carrier (5 kohms) |
|---|---|---|
| Latency (µs) 20% of signal | 2 µs ( 50% Ge) | Higher with 75% Ge |
| Pixel dimensions(µm x µm) | 1 µm X 0.1 µm | Down-scalable |
| Buried gate thickness (nm) | 20 nm | Can be reduced to 10 nm |

## 2. Readout strategies and operation mode

The readout can be simply made by switching on the single (1T) n-channel- TRAMOSFET. The source to drain current magnitude simply contains the information (pixel hit or not hit). In addition the magnitude of the current signal is directly related to the number of charges generated in the sensitive volume of the pixel. Hence this could provide additional extra spectroscopic data adding to the point of impact information this 1T pixel detector can achieve. In the case of X ray photons this property should be considered as useful given the spectroscopic information that is not directly obtained on simple position sensitive detectors. In this case the pixel should be operated in full analog/digital mode with a multi-bit resolution readout.

One item to be regarded as important is the fact that the pixel does not dissipate any power in the detection mode. Only during readout power is dissipated with an energy budget of the order of 5x10 pJ, if we can reach a 10 ns limit to single pixel readout duration, as TCAD simulations suggest.

The pixel is designed with collecting lengths not exceeding the 10 nm limit. This means that with a carrier velocity of $10^7$ cm.s$^{-1}$ the collecting time is of the order of $10^{-6}/10^7$ equal to $10^{-13}$ s = 100 fs, this means that the with a lifetime of less than 1 ns the charge collection should be complete, this has been checked by device simulation. Because of these effects that are purely geometrical the device should be less sensitive to bulk defects (point like or extended) induced by massive particles. As a reminder, at room temperature (and in germanium) neutron induced defect are localized in regions of 1 µm in size [6-7]. The possibility of damaging a single pixel by one interacting particle (NIEL) should then be considered, with a generation of a "cold" or "hot" pixel.

### 2.1. Front end readout

In a pixel array (organized as a position sensitive detector) ,binary readout with a predefined threshold is sufficient to obtain a micrometer range resolution. It was done previously with a CIS based array which exhibit for the first time a sub 5 µm point to point resolution using binary readout of 25 µm x 25 µm pixels [8]. For the same purpose there are many readout strategies. First the analog one with comparators located at bottom column level, giving a binary (hit/no-hit) output which is in many respects similar to the one used in [8]. Second one where the comparator/latch is located on the pixel. This was proposed in 2011 (Sensors 2011, [9]). A similar circuit that can be used for this purpose is described in Fig. 5.

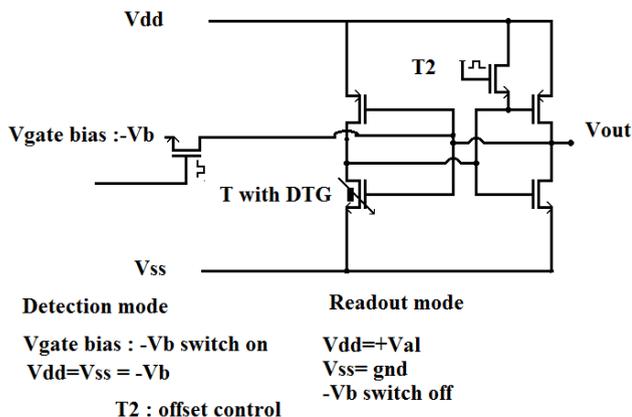

Fig. 5: A proposed binary on pixel readout which uses 4 extra transistors. With a sub-micron process (< 50 nm technological node) this would be possible with a limited space budget (< 0.1 µm² area budget).

The first readout technique is still appropriate in the use of a single transistor pixel. Because of the bias scheme in readout and detection mode, only a one column line is necessary for each column, the signal being discriminated at column level just as in CIS readout schemes. Parallel column readout is made, at the bottom of the array. The data can then be processed to reduce the flux that must be serialized and exported out of the pixel array monolithic chip.

## 2.2. Data compression and serial output: a reminder

In a practical high energy particle physics experiment, the number of pixels hit per unit time is proportional to the event rate ate the interaction point which is itself proportional to the luminosity (cm²s⁻¹). This does not depend on the number of pixels per unit area, thus with a high density of pixels of small area per pixel, the number of hit pixels should remain constant (per unit time). Simple estimations can be made on the variation of the data flux. Hence, for a given area he number hits during a time interval is proportional to the event rate at the interaction point. This event rate is proportionally dependent on the luminosity and hence the data (hit/no hit) does not depend on the number of pixels per unit area. However the pixel-address length depends on the number of pixels in a given area as. For a given unit area let $N^2$ be the number of pixels in the corresponding squared array. For an address length of n bits, $2^n$ is the number of pixels. With all calculations made, n varies proportionally to log (N). Thus the data flow will be proportional to the log (N) x Event rate. This means that many zero hit pixels are numerous and that a zero suppression scheme is necessary on each array. Such schemes have been implemented on prototypes for experiments, and can be transferred into the modified CMOS process that will be use for the TRAMOS pixel.

## 2.3. Alternative pixel architectures and operational modes.

This pixel concept has been fully simulated using an n-channel based device. The device has two operation states/ detection and readout. For fast detection and time tagging the pixel must be in a same detection/readout mode. With the QW in the VB this means that the p-channel transistor should be used. With a negatively biased gate it will see a constant source-drain current changed when e-h are generated in the depletion zone and following that the holes become localized in the QW acting as a buried gate. If the device is fast enough to switch on the pixel can charge a one-pixel capacitor (at constant current) hence giving an analog signal proportional to the time elapsed since the hit. This requires somehow a permanent power dissipation in the pixel. To have clear view if the power per unit area (cm²) should not exceed 1 W for a 1μm² pixel it should not exceed $10^{-8}$ W=10 nW so with a 3 V supply voltage the current should be lower than 3 nA, which means that the device leakage current should be drastically reduced and that the 10 nm node technologies should be chosen. This is only an average value and with the circuit proposed in Fig. 5 this could be within the reach of current technologies, with thermal dissipation only being significant after a hit. An on-pixel capacitor of 1fF supplied with a 10 nA current source would provide a ramp dV/dt = $10^{-9+15}=10^6$ V/s or 1 mV/ns. Of course it is difficult to estimate a timing resolution from such a design, it could be better than 1 ns, allowing some time tagging to be implemented as well as position information.

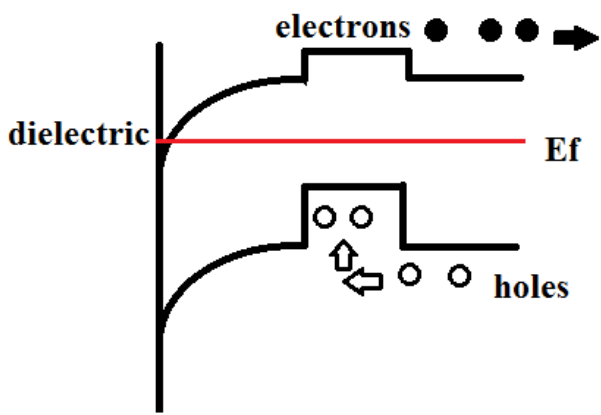

Fig. 6: Band-diagram of a proposed p-channel transistor comprising a QW and bias in the "on" mode with a negative voltage on the upper gate (not in the figure). For simplicity the Fermi-level or quasi Fermi-level is shown as flat. Holes should migrate (by drift) towards the QW and electrons towards the substrate. Readout and detection are the same.

### 2.4. Single transistor FDSOI pixel

As a backup for the TRAMOS pixel is the single transistor FDSOI pixel. Up to now multi-transistor pixels with direct coupling have been studied and some capacitive coupling has been proposed. The biasing scheme is similar to the one used in the basic TRAMOS device. The only difference being that in this case the bulk of the Silicon On Insulator substrate is used as a detecting volume and a collecting electrode should be introduced below the $SiO_2$. Current FDSOI have a BOX (Buried OXide) thickness of 20 nm (T. Ernst LETI, Private Communication). Thus a capacitive coupling can be made using the BOX and a back gate effect can be effective to modulate the source-drain current. This was considered for SOI/JFETs [10] of the DMILL [11] technology back to the nineteen nineties but the buried oxide then had a thickness too high to make an effective coupling. Direct coupling to the upper was then considered but this was consequently considered by other groups. FDSOI like devices can be used either in the electron or holes mode. Some simulation work is now necessary to figure out some quantities such as the CVF (conversion factor).

### 3. Process a TRAMOS : building blocks :

Some ongoing work has started a year ago to study the process stages necessary for the fabrication of these pixels arrays. First ion implantation is used (at high energy) to create buried layers that can be characterized using SIMS (Secondary Ion Mass Spectroscopy) or RBS (Rutherford Backscattering). Additionally room temperature Raman scattering was made on the some samples. Table 2 summarizes the results. SIMS results were obtained at the peak concentration whereas RBS results were obtained using a fit. The region above 600 nm towards the surface is almost depleted of Ge. This could then explain the lower values obtained using RBS for the profile determination. If the concentration determined by SIMS is integrated along the implanted zone we obtain a value slightly exceeding the total flux received by the sample ($5 \times 10^{17}$ cm$^{-2}$ instead of $2.3 \times 10^{17}$ cm$^{-2}$). This discrepancy being certainly due to a background found in the SIMS profile that could be not physical.

Table 2: significant results of the 1 MeV Ge implantation, using SIMS and RBS.

| $^{72}$Ge ion | SIMS (Cs+) positive secondaries) | RBS (using a fit program) |
|---|---|---|
| Sample C1 | $1.35 \times 10^{22}$ cm$^{-3}$ (30%) | |
| Sample C2 | $1.35 \times 10^{22}$ cm$^{-3}$ (30%) | |
| Sample D1 | | $3.25 \times 10^{21}$ cm$^{-3}$ (7%) |
| Sample D2 | | $4.50 \times 10^{21}$ cm$^{-3}$ (9%) |

The target 25% Ge in Si is attained. However with no post-implantation annealing Raman scattering measurements in Stokes mode show that region below the surface (a few 100 nm) contains many defects. Most of the Silicon lines vanish and the remaining lines being noisy and not characteristic of crystalline silicon. After high temperature anneal (850 °C during 50 minutes) the Raman spectrum recovers into a typical Si spectrum with some lines originating from Si-Ge or Ge-Ge modes appearing. For the fabrication of the deep n-well some Phosphorous implants have been made with results summarized in Table 3:

Table 3: Peak Phosphorous concentration after 14 MeV P ion implantation

| Sample | Peak | Background | Integral | Peak (μm) |
|---|---|---|---|---|
| A1W24004 (BR) | $5 \times 10^{16}$ cm$^{-3}$ | $10^{12}$ cm$^{-3}$ | 3.38e12cm-2 | 6 μm |
| A2W2404 (BR) | $6 \times 10^{16}$ cm$^{-3}$ | $10^{12}$ cm$^{-3}$ | 4.16e12cm-2 | 6 μm |
| A3W2404 (BR) | 7x10$^{16}$cm$^{-3}$ | 10$^{12}$cm$^{-3}$ | 4.19e12cm-2 | 6 µm |
| SRIM simulation | 4x10$^{16}$cm$^{-3}$ | 10$^{12}$cm$^{-3}$ | | 6.18 µm |

This shows that this technique allow the fabrication of a buried n-well, with a high energy implantation without damaging the material above the n-well. A deep n-well is useful to reduce leakage current through the substrate. These process test will be followed by test of UHV/CVD SiGe layer fabrication and characterization. Up to now no critical problems and bottlenecks have arisen.

## 4. Concluding remarks

It is far too early to draw definitive conclusions about the outcome of this R&D. The soundness of the designs presented here has been assessed by extensive simulations, although more simulation work should be made on the newly proposed p-channel TRAMOS cell. The technological steps needed to make a fabrication process flow are now being studied by simulations, analytical calculations and computations using physical quantities from recent published experimental results. Measurements on realistic test vehicles are now under way focusing on two techniques: ion implantation and UHV/CVD layer growth. A CMOS compatible process with a SiGe add on fabrication step will be a successful outcome of this R&D. One subsequent step will be to assure the radiation hardening of the process but this seems to our reach for the present-day technological nodes [6][12-13].

**Acknowledgements:**

The authors would like to thank P.F. Honoré (I.T. Information Technologies) and F. Eozenou (chemical processing for some samples) for some preliminary work. High energy ion implantation were made at Jannus/EMIR facility and we acknowledge the assistance of the staff. RBS measurements were made at CSNSM by Cyril Bachelet and co-workers. Details about UHV/CVD were provided by Charles Renard and Geraldine Hallais (C2N, Orsay) where some high temperature annealing were performed.